\documentclass[twocolumn,showpacs,amsmath,amssymb,pra,nofootinbib]{revtex4}

\usepackage{graphicx}
\usepackage{dcolumn}
\usepackage{bm}

\newcommand{\be}{\begin{equation}}
\newcommand{\ee}{\end{equation}}
\newcommand{\bea}{\begin{eqnarray}}
\newcommand{\eea}{\end{eqnarray}}

\newcommand{\ra}{\rightarrow}

\newcommand{\RR}{\rangle}
\newcommand{\LL}{\langle}

\newcommand{\RL}{\RR\LL}

\begin{document}
\bibliographystyle{apsrev}

\title{Entanglement Dynamics in 1D Quantum Cellular Automata\\}

\author{Gavin K. Brennen and Jamie E. Williams}
\address{National Institute of Standards and Technology, Gaithersburg, Maryland 20899-8423\\
}

\date{\today}

\begin{abstract}
Several proposed schemes for the physical realization of a  quantum computer consist of qubits arranged in a cellular array. In the quantum circuit model of quantum computation, an often complex series of two-qubit gate operations is required  between arbitrarily distant pairs of lattice qubits.  An alternative model of quantum computation based on quantum  cellular automata (QCA) requires only homogeneous local interactions that can be implemented in parallel. This would  be a huge simplification in an actual experiment. We find some minimal physical requirements for the construction of unitary QCA in a 1 dimensional Ising spin chain and demonstrate optimal pulse sequences for information transport and entanglement distribution.  We also introduce the theory of non-unitary QCA and show by example that non-unitary rules can generate environment assisted entanglement.
\end{abstract}


\maketitle

\section{Introduction}
Much progress has been made recently in developing architectures that can support quantum information processing (QIP).  The key result on the universality of quantum computers (QC) is that given the ability to implement single qubit and two-qubit gates in a network of connected qubits any computation or quantum simulation can be implemented.  However, some systems are better adapted to implement certain QIP tasks than others.  In particular, lattice based systems with regularly arranged qubits interacting with nearest neighbors such as a neutral atom optical lattice, quantum dot arrays, and Phosphorus embedded Silicon offer several advantages in terms of reconfigurability and scalability.  For example, optical lattices with nearest neighbor tunneling couplings have been shown to be a promising platform to simulate many body Hamiltonians \cite{Duan:Oplatsim}.  Generally, lattices are well suited to perform parallel computation protocols such as entanglement distribution \cite{Briegel:1wayQC} and entanglement swapping \cite{Brennen:Qarch}.

The natural facility of these systems invites study of other models of computation that take advantage of the lattice architecture.  Perhaps the most relevant computational model in classical systems is a cellular automaton.  The essential idea behind cellular automata (CA) is to make use of simple local rules uniformly applied across a lattice of cells to generate complex dynamics. Depending on the initial state of the system and the underlying rule, long range spatial and temporal correlations can develop resulting in complex behavior.  Classical CA can simulate a wide range of complex physical phenomena including fluid dynamics, nonlinear diffusion, percolation, and phase transitions in many body systems \cite{Wolfram:NKS}.  Formally, a CA is termed complex if it evolves in a manner that in some sense is computationally irreducible, meaning it cannot be predicted with a compactly written equation \cite{Crutchfield:1994a}.  A number of CA rules have been shown to be computationally universal, in the sense that they can emulate a universal Turing machine \cite{Wolfram:NKS}.

The extension of the cellular automaton concept to quantum systems is fairly straightforward, though
as we will show, requires a slight modification of the classical CA procedure for implementing the local rules. For a two-state CA, which will be the focus of this paper, each cell in a quantum cellular automata (QCA) corresponds to a qubit that can be in a superposition of states $|0\rangle$ and $|1\rangle$ and the local rule is carried out via a unitary gate operation on each neighborhood. The essential new feature in a QCA that makes it distinct from its classical counterpart is that nonlocal correlations can develop between cells resulting in the spread of entanglement throughout the system. This property of QCA will be of central importance in this paper.

Our motivation for studying QCA is to explore the power of low computational depth circuits applied in uniform across a system to produce complex quantum dynamics. This is in marked contrast to the typical QC approach, where a complex sequence of logic gates acting on distributed qubits in the computer is carried out in a serial fashion in order to produce the desired output of a specific computation. Most previous work on QCA has focused on mapping such systems to the QC circuit model \cite{Watrous:QCA,Benjamin:QC4cellen}.  Additionally, there have been investigations of quantum lattice gas automata (QLG) for simulations of the Dirac equation in 1D \cite{Meyer:QLG1} and for topological computation \cite{Meyer:QLG2}.  Recently there was an experimental realization in liquid state NMR of a QLG algorithm to solve the 1D diffusion equation \cite{Pravia:QLG}.  We propose using 1D QCA to explore complex quantum correlations generated by simple rules applied over small neighborhoods.  Characterizing multi-particle entanglement is a field of active research both for itÕs potential use in QIP and in the study of non-locality in physics.  QCA can offer a unique approach to study the raw computational effort needed to generate such entanglement.

From an experimental standpoint, a QCA has a significant advantage over a QC because individual qubits in the lattice do not need to be separately addressed, since uniform rules are applied in parallel across the lattice. In such an implementation, applying uniform fields over the entire system helps to eliminate error resulting from Òcross talkÓ on neighboring qubits due to imperfectly aligned control fields.  Some specific physical systems have been proposed as candidates for QCA including quantum dot arrays \cite{Toth:QdotQCA} and endohedral fullerenes \cite{Twamley:QCAQCA}. Throughout the development of the general QCA formalism in this paper, we provide specific examples of possible experimental implementations in order to emphasize the relevance of the QCA approach to present day technologies.

In Sec. II we introduce the formalism for QCA and show how to construct arbitrary three cell neighborhood rules using homogeneous pairwise interactions and single qubit gates.  We show how to transport quantum information with QCA in Sec. III and demonstrate optimal sequences to swap two distant quantum states and to prepare three types of entangled states.  In Sec. IV we explore general properties of entanglement dynamics with QCA.  The dynamics of multi-spin entanglement are measured by a function linearly related to the purity of the single qubits averaged over the lattice.  This measure has the advantage of being observable in a physical system supporting the QCA architecture.  In Sec. V we extend the theory to open systems and demonstrate how more general non-unitary rules can be implemented in the QCA paradigm using measurement and quantum feedback.  It is shown that for a particular mixing of a non-unitary rule with a unitary rule, entanglement is generated across a spin chain where there is none for purely unitary evolution.  This is an example of environment assisted entanglement generation.  Finally, we present conclusions and open questions in Sec. VI.
\section{Formalism}\label{sec:Formalism}
\subsection{Simulating QCA rules}\label{sec:Simulating}
Consider a 1D array of $n$ lattice sites occupied by qubits ordered $0$ to $n-1$.  We define a radius $r$ QCA as one that changes the state of a qubit at site $j$ dependent on the states of the qubits in the neighborhood $[j-r,j+r]$.  Given a system with nearest neighbor interactions, the simplest unitary QCA rule has $r=1$ describing a unitary operator applied over a three cell neighborhood $(j-1,j,j+1)$:
\be
\begin{array}{lll}
M(u_{00},u_{01},u_{10},u_{11})&=&|00\RR\LL00|\otimes u_{00} +|01\RR\LL01|\otimes u_{01}\\
&+&|10\RR\LL10|\otimes u_{10}+|11\RR\LL11|\otimes u_{11}.\;
\end{array}
\label{QCArule}
\ee
where $|ab \RR\LL ab|\otimes u_{ab}$ means update the qubit at site $j$ with the unitary $u_{ab}$ if the qubit at site $j-1$ is in state $|a\RR$ and the qubit at site $j+1$ is in state $|b\RR$.  In classical CA the local update rule $M$ can be applied in parallel to all cells.  To do so requires that a separate register store the current state of the lattice so the previous state of the neighbors is known before the cells are updated in parallel.  For instance, radius $1$ CA rules could be implemented by copying the current state, updating the even ordered cells on the original and the odd ordered cells on the copy, and splicing the updated cells together.  By the no cloning theorem \cite{Wooters:NoCloning}, non-orthogonal quantum states cannot be copied so this is not possible for QCA. 
However, the update can be divided into two stages:  first update all the even qubits with rule $M$, next update all odd qubits.  This rule is denoted a Block partitioned QCA (BQCA) and guarantees that at each stage the operators commute and thus can be implemented in parallel \cite{Wolfram:NKS}.

We show that any BQCA can be simulated with a lattice of even order constructed with an alternating array of two distinguishable species $ABABAB\ldots$ that are globally addressable and interact via the Ising interaction.  In deriving the construction of QCA rules we initially assume periodic boundary conditions ($n+j\equiv j$).  The simulation is shown to be easily adapted to a lattice with fixed boundaries. 

The general pairwise interaction Hamiltonian across a 1D lattice is
\be
H_{I}(t)=\sum_{j=0}^{n-1}\sum_{\alpha,\beta=0}^{3} g_{\alpha,\beta}^{j}(t)  \sigma_{\alpha}^{j}\otimes \sigma_{\beta}^{j+1},\;
\ee
where the Pauli operators are labelled $\{\sigma_{\alpha}\}=\{1,\sigma_{x},\sigma_{y},\sigma_{z}\}$ and the $g_{\alpha,\beta}^{j}(t)$ are, possibly time dependent, coupling strengths ($\hbar=1$).  It is straightforward to show that in order for the Hamiltonian to commute over all nearest neighbor pairs with periodic boundary conditions, it is restricted to the form
\be
H_{I}(t)=\sum_{j=0}^{n-1} g^{j}(t)\sigma_{\vec{r}_{j}}^{j}\otimes \sigma_{\vec{r}_{j+1}}^{j+1}+\sum_{j=0}^{n-1}h^{j}(t)\sigma_{\vec{r}_{j}}^{j},\;
\label{sums}
\ee
where $\sigma_{\vec{r}_{j}}^{j}\equiv\vec{\sigma}\cdot\vec{r}_{j}$ defines the local Bloch vector at site $j$.  We identify the local basis of each qubit along this Bloch vector meaning $\sigma_{z}^{j}\equiv\sigma_{\vec{r}_{j}}^{j}$.  The second sum in Eq. \ref{sums} refers to single qubit free Hamiltonians.  Note that in order to satisfy the periodic boundary conditions, $n$ must be even.

The system dynamics can be controlled in a non-trivial way with limited addressability by assuming a 1D lattice constructed two species of qubits $A$ and $B$ arranged in antiferromagnetic order.  Here the species may have distinguishable two level energy spacings, $h^{j}=h^{A(B)}$ for $j$ even(odd), meaning the species are addressable in frequency allowing  even or odd ordered qubits to interact in parallel with an external field.  The two species could also correspond to disjoint two dimensional subspaces of the same four dimension system.  In either case, a general control Hamiltonian that performs single qubit rotations on the two species is written:
\be
H_{C}(t)=\vec{\Omega}_{A}(t)\cdot\sum_{j=0}^{n/2-1}\vec{\sigma}^{2j}+\vec{\Omega}_{B}(t)\cdot\sum_{j=0}^{n/2-1}\vec{\sigma}^{2j+1}.\;
\ee
The total Hamiltonian acting on the system is $H=H_{I}(t)+H_{C}(t)$.  For simplicity, we assume an isotropic pairwise interaction $g^{j}(t)=g(t)$ corresponding to the Ising interaction, and transform $H$ to the appropriate rotating frame so that the total Hamiltonian becomes:  
\be
H^{\prime}(t)=H_{I}^{\prime}(t)+H_{C}^{\prime}(t)=g(t)\sum_{j=0}^{n-1} \sigma_{z}^{j}\otimes\sigma_{z}^{j+1}+H_{C}^{\prime}(t).\;
\ee
Discrete time dynamics describing cellular automata can be implemented with continuous dynamics by first evolving the system with the interaction Hamiltonian followed by evolution by the control that performs simultaneous single qubit gates on either or both species.  This can be realized with a fixed Ising interaction punctuated by ``hard" control pulses as is done in NMR pulse sequencing \cite{Ernst:QCont}, or the physical system may allow the pairwise couplings to be turned off during the single qubit gates.  In any case, because the interaction Hamiltonian commutes with itself at all times, the unitary corresponding to coupled evolution can be written $U(t)=e^{-i\int_{0}^{t}H_{I}^{\prime}(t)dt}$ and the single qubit gates are generated by $H_{C}$.  

The simplest non trivial BQCA rule obtainable in the two species architecture is described by the following gate sequence:
\be
M({\bf 1},u,u,u^{2})=e^{-i\frac{\pi}{2}\sigma_{\vec{m}}}U([\gamma/(2 g)])e^{i\gamma\sigma_{z}}e^{i\frac{\pi}{2}\sigma_{\vec{m}}},\;
\label{Mtot}
\ee
where $u=e^{i \gamma \sigma_{\vec{n}}}$ is an arbitrary element of SU(2) written as a rotation about the Bloch vector $\vec{n}=(\sin{\theta}\cos{\phi},\sin{\theta}\cos{\phi},\cos{\theta})$, and $\vec{m}=(\sin{\theta/2}\cos{\phi},\sin{\theta/2}\cos{\phi},\cos{\theta/2})$.  The evolution time in brackets, viz. $U(t=[x])$, means add the appropriate multiple of $\pi/|g|$ to  the quantity $x$ to make it positive.  Henceforth, we assume $g>0$.   All qubits of species $A(B)$ are updated by the rule $M^{A(B)}$ when the single qubit gates act on that species.  A single step of the BQCA is defined as the sequence:  $M\equiv M^{A}M^{B}$ which updates all cells, and the BQCA rule iterated $t$ times on the initial state $|\psi(0)\RR$ generates the state $|\psi(t)\RR=[M]^{t}|\psi(0)\RR$.  

The particular rule Eq. (\ref{Mtot}) is left/right symmetric with the interpretation that the center qubit is rotated by an amount proportional to the sum of the qubit values of the neighbors.  Note that $M^{A}(u^{2},u,u,{\bf 1})=\sigma_{x}^{B}M^{A}({\bf 1},u,u,u^{2})\sigma_{x}^{B}$.
Another elementary rule is:
\be
\begin{array}{lll}
M({\bf 1},u,u,{\bf 1})&=&-e^{-i\frac{\pi}{2}\sigma_{\vec{m}}}[e^{-i\frac{\pi}{2}\sigma_{y}}U([\gamma/(4 g)])e^{i(\frac{\pi}{2}-\frac{\gamma}{2})\sigma_{z}}\\
& &e^{i\frac{\pi}{4}\sigma_{y}}
U([-3\pi/(4 g)])e^{i \frac{3\pi}{2}\sigma_{z}}e^{-i\frac{\pi}{4}\sigma_{z}}e^{i\frac{\gamma}{2}\sigma_{y}}\\
& &U([-\pi/(4 g)])e^{i \frac{\pi}{4}\sigma_{y}}e^{i\frac{\pi}{4}\sigma_{z}}e^{i (\frac{\pi}{4}-\frac{\gamma}{2})\sigma_{x}}\\
& &e^{i\frac{\pi}{2}\sigma_{z}}e^{i\frac{\pi}{4}\sigma_{y}}U([\gamma/(4 g)])e^{-i\frac{ \gamma}{2}\sigma_{z}}]e^{i\frac{\pi}{2}\sigma_{\vec{m}}},\;
\end{array}
\label{M1}
\ee
where the unitary $u$ and the Bloch vectors $(\vec{n},\vec{m})$ are defined as above.  Combining rules \ref{Mtot} and \ref{M1} we can construct all symmetric QCA rules:
\be
\begin{array}{lll}
M(u_{00},u_{01}=u_{10},u_{11})&=&M(v^{2},v,v,{\bf 1})M({\bf 1},w,w,{\bf 1})\\
& &M({\bf 1},u,u,u^{2}),\;
\end{array}
\label{symrule}
\ee
where $v=u_{00}^{1/2}, u=u_{11}^{1/2}$, and $w=u_{00}^{-1/2}u_{01}u_{11}^{-1/2}$.  A maximum of six pairwise interactions $U$ interspersed by single qubit gates is sufficient to simulate the symmetric rules.

Asymmetric rules can be constructed if the Ising interaction is allowed to have different coupling strengths between left-center and center-right pairs.  The appropriate Hamiltonian is:
\be
H_{asym}=\sum_{j=0}^{(n-2)/2}({g^{1}(t){\sigma_{z}^{2j}\otimes\sigma_{z}^{2j+1}+g^{2}(t)\sigma_{z}^{2j+1}
\otimes\sigma_{z}^{2j+2}}}).\;
\ee
This asymmetry can be built into the system as is suggested, for instance, in proposals to implement quantum computation in optical lattices \cite{Brennen:QC1}.  Here atoms are trapped in a 3D periodic potential created by standing waves of interfering laser beams and prepared with one atom per potential well.  An antiferromagnetic ordering of atomic species can be created along one dimension, and by appropriate tuning of the laser parameters, wells can be joined along this dimension such than each atoms interacts with its left or right neighbor.  By choosing different interaction strengths (or times) between the center-left and right neighbors, $H_{asym}$ can be simulated.  

Given the ability to implement $H_{asym}$, the following rule can be generated:
\be
M({\bf 1},u^{-1},u,{\bf 1})=e^{-i\frac{\pi}{2}\sigma_{\vec{m}}}U([t])e^{i\frac{\pi}{2}\sigma_{\vec{m}}},\;
\ee
where the time and couplings satisfy
\be
\begin{array}{lll}
\int_{0}^{t}(g^{1}(t)-g^{2}(t))dt&=&\gamma,\\
\int_{0}^{t}(g^{1}(t)+g^{2}(t))dt&=&0\pmod{2\pi},\;
\end{array}
\ee
and $(u,\vec{m},\vec{n})$ are as above.  A general QCA rule can be constructed from the above elementary rules:
\be
\begin{array}{lll}
M(u_{00},u_{01},u_{10},u_{11})&=&M({\bf 1},x^{-1},x,{\bf 1})M({\bf 1},x,x,{\bf 1})\\
& &M(v^{2},v,v,{\bf 1})M({\bf 1},w,w,{\bf 1})\\
& &M({\bf 1},u,u,u^{2}),\;
\end{array}
\ee
where $v=u_{00}^{1/2},x=\sqrt{u_{10}u_{01}^{-1}},w=u_{00}^{-1/2}u_{01}u_{11}^{-1/2}$, and $u=u_{11}^{1/2}$.  A maximum of 11 sequences of left/right shifts punctuated by single qubit gates are sufficient to implement an arbitrary 3 cell BQCA rule $M$, although it is uncertain whether this is optimal.

The present construction of $M(u_{00},u_{01},u_{10},u_{11})$ with $u_{ij}\in$ SU(2) is only a subset of the most general rule having three additional relative phases, $M(u_{00},e^{i\phi_{1}}u_{01},e^{i\phi_{2}}u_{10},e^{i\phi_{3}}u_{11})$.  One relative phase can be fixed by applying a $\hat z$ rotation, $e^{i \frac{\phi_{1}}{2} \sigma_{z}}$, to the neighboring qubits after an update.  Also, for each unitary $u=e^{i \gamma \sigma_{\vec{n}}}$ the replacement $\gamma\rightarrow\gamma+\pi$ introduces a sign change.  A second relative phase is fixed by applying different controlled phase gates between a qubit and each neighbor (as is possible with the interaction $H_{asym}$).  Generating a third relative phase requires a direct interaction between left and right neighbors, not possible in the 1D architecture with only nearest neighbor connectivity and two species addressability.           
\subsection{Boundary Conditions}
In the above treatment we have assumed periodic boundary conditions, in which case the the BQCA rules can be implemented uniformly with only global addressability of two species.  In practice, a system coupled by the Ising interaction with periodic boundaries could be realized by a custom designed ring molecule with alternating atomic species, or perhaps with trapped atoms in a ring type cavity.  In most experimental situations it will be easier to construct a linear system with boundaries.  Consider an open 1D spin chain labeled from left to right by the integers $0$ to $n-1$, where as above $n$ is assumed even.  One can simulate evolution where each cell is updated according to neighborhood values by introducing fictitious boundaries on the left and right ends with fixed values:  $\Sigma_{L,R}\in\{0,1\}$.     This is accomplished by appending appropriate single qubit gates to the ends of the chain after each instance of $U(t)$ in the above rules.  For a sequence updating even ordered $(A)$ species, append $U(t)$ with $e^{-i(-1)^{\Sigma_{L}}gt\sigma_{z}^{0}}$, and for a sequence updating odd ordered $(B)$ species, append with $e^{-i(-1)^{\Sigma_{R}}gt\sigma_{z}^{n-1}}$. In this way, addressability at the ends (or at least the ability to introduce energy shifts at the ends) is sufficient to simulate BQCA rules over 1D systems with boundaries.
\subsection{Universality}
We have identified a finite set of rules to construct a class of radius 1, two state BQCA.  We now discuss some issues regarding the universality of this class of quantum cellular automata.  By universality we refer to the ability to emulate other computations, in particular other QCA and quantum computers, in an efficient manner.  A distinguishing feature of computation with classical cellular automata is that CA have minimal time complexity in that the same rule is applied to the data register at each iteration.  This is in contrast to the conventional computers that use a complex sequence of logic gates over the period of computation.   It has been shown that a radius 1, two state classical CA rule, designated rule 110, is universal in the sense that by appropriate choice of initial state it can emulate any other CA as well as a turing machine \cite{Wolfram:NKS}.  It should be emphasized that this rule updates all cells synchronously.  It is not obvious that by appropriate choice of initial state, a single BQCA rule would be universal in the same way.  It can be shown, however, that a sequence of rules can simulate a quantum computer with only linear cost in space and time resources.  This may violate the spirit of using a single rule to generate complex dynamics but demonstrates that the underlying physical architecture for QCA supports universal computation.  The ability to simulate a QC follows from the work of Benjamin who has shown \cite{Benjamin:QC1cellen,Benjamin:QC4cellen} that an open 1D lattice composed of an alternating array of two species of qubits can be used for quantum computation.  The only architectural requirements are global addressability of the species and addressable boundaries.  He proposes two protocols with different physical assumptions.  

The first protocol \cite{Benjamin:QC4cellen} assumes that unitaries $S^{u}_{f}$ can be implemented in parallel,  meaning ``apply the unitary $u$ to species $S$ if the field value is equal to $f$".  The field value is defined as the number of 0's minus the number of 1's in the neighborhood of each qubit, therefore $f\in\{-2,0,2\}$ inside the lattice and $f\in\{-1,1\}$ at the boundaries.  
BQCA rules provide an explicit construction of these unitaries, viz. the sequence $S^{u}_{2}S^{v}_{0}S^{w}_{-2}$ acting on interior spins is exactly simulated by the symmetric BQCA rule $M^{s}(u,v,v,w)$.  In this proposal the boundary spins need to be addressable in order to load information into the lattice, but the only operations needed are the bit flips, $S^{\sigma_{x}}_{-1,1}$.  When the entire lattice is coupled via the Ising interaction, this is achievable by dynamically decoupling the boundary spins from the rest of the lattice using standard refocusing techniques \cite{Ernst:QCont}.  Explicit pulse sequences to perform computation with endohedral fullerenes in a QCA architecture have been recently proposed in \cite{Twamley:QCAQCA}.  

In the second protocol \cite{Benjamin:QC1cellen} it is not necessary to apply unitaries that depend on the total field value, however, it is assumed that the Hamiltonian between left/center neighbors, $H^{AB}$, can be turned ``off" while the center/right Hamiltonian, $H^{BA}$, is turned ``on", and vice versa.  This is akin to the physical requirement for asymmetric BQCA rules and may be more difficult to engineer in a given system.  

The ability to map BQCA to Benjamin's model of computation resolves a question about whether BQCA are universal with respect to the ability to efficiently simulate other quantum cellular automata.  Watrous \cite{Watrous:QCA} has shown that 1D-partitioned QCA can be simulated by a
quantum turing machine QTM) with only linear slowdown.  1D-partitioned QCA are a restricted class of
1D-QCA in which each cell is partitioned into three subcells and the rule
updates the cells by permuting subcells of neighboring cells and operating on
the new cells in parallel with quasi-local unitary operations.  van Dam
\cite{vanDam:QCAth} extended this result to prove that quantum gate cellular automata (QGCA)
can simulate any unitary QCA with only a polynomial slowdown.  QGCA evolve
by a repeated sequence of two steps: one step acts to permute
the basis states within a certain neighborhood, and the second step applies
parallel quasi-local gates over the neighborhood.  Not all of the quasi-local gates in the QGCA model can be implemented with pairwise interactions and two species addressability.  However, one can use Benjamin's protocols to show that with a properly prepared initial state, a sequence of homogeneous update rules can simulate the QGCA model with only a linear cost in space and time resources.  Because BQCA rules are sufficient to implement Benjamin's model, BQCA are also universal in this respect.   

\section{Information Transport}\label{Sec:inftrans}
The discrete time process corresponding to QCA evolution is a useful way to study information flow in quantum systems.  For classical CA the maximum speed of information flow, $c_{max}$, is 1 cell per update, which defines a light cone for information propagation.  This can be realized, for instance, by beginning in the state $0\ldots010\ldots0$ and evolving with Wolfram's rule 254 \cite{Wolfram:NKS}.  This rule updates the center cell in a three cell neighborhood, mapping each cell to a $1$ unless its left and right neighbors are in state $0,0$.  Evolving the initial state will cause the string of $1$'s to grow by one cell on the left and right at each step.  This does not fit into the QCA paradigm for two reasons.  Firstly, the local rule is not unitary, e.g. both strings ``110" and ``111" are mapped to ``111".  Secondly, the applied rule updates all cells simultaneously, not in a block partitioned manner.  There are locally reversible CA rules that spread information at speed $c_{max}$ (such as rule 150), but is there a unitary BQCA rule that can saturate the speed limit?  The answer is affirmative as is shown below.       

We consider an $n$ cell register initialized in the state $|1\RR_{0}\otimes|0\RR_{1\ldots n-1}$ with left and right boundary conditions $(\Sigma_{L}=0,\Sigma_{R}=0)$.  The approach is to map the solitary $1$ into a two cell unit which then propagates 2 sites per update and is decoded into a single cell at the other boundary.  The BQCA sequence to achieve this is:  $\sigma_{z}^{n-1}[M({\bf 1},e^{-i\frac{\pi}{2}\sigma_{x}},e^{-i\frac{\pi}{2}\sigma_{x}},e^{-i\pi\sigma_{x}})]^{n/2}$.   The total time to transport the information over $n-1$ cells is (from Eq. \ref{Mtot}) $T_{n}=n\pi/(4g)+(n+1)t_{s}$, where $t_{s}$ is the time to implement a single qubit gate.  If we assume that single qubit gates can be implemented on a time scale much faster than the many body interaction, $t_{s}\ll\pi/(4 g)$, then the information speed is $c=(n-1)/T\simeq4g(n-1)/(\pi n)$.  The connection is made to classical CA's by noting that nontrivial rules use conditional bit flips , which, according to Eq. \ref{Mtot} are implemented in a time $t=\pi/(4g)$.  Thus the maximum speed of information flow is $c_{max}=1/t=4g/\pi$.  The encoding and decoding consume a fixed amount of time but in the limit of large $n$, $c_{max}$ is approached.   

\begin{figure}
\begin{center}
\includegraphics[scale=0.5]{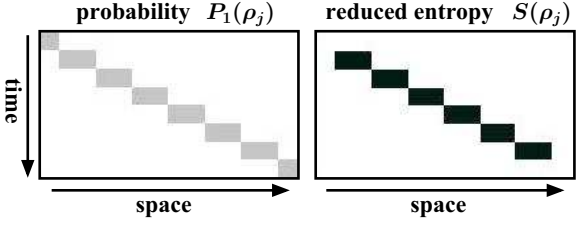}
\caption{\label{fig:1}Transporting a quantum state over an $n=14$ cell 1D lattice via BQCA evolution by the rule $M({\bf 1},e^{-i\frac{\pi}{2}\sigma_{x}},e^{-i\frac{\pi}{2}\sigma_{x}},e^{-i\pi\sigma_{x}})$.  Space time diagrams are shown with cell number on the horizontal axis and time flowing downward.  On the left is a history of cell site probability to be in state $|1\RR$ and on the right, the reduced von Neumann entropy of each cell (Black=1,White=0).}
\end{center}
\end{figure}

\begin{figure}
\begin{center}
\includegraphics[scale=0.5]{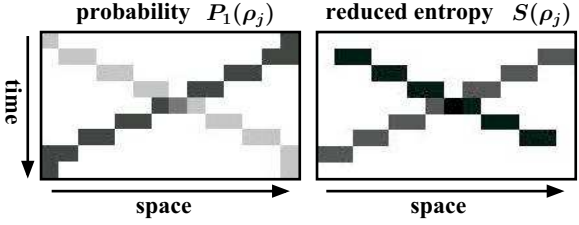}
\caption{\label{fig:2}Swapping the states of cells at the boundaries of an $n=14$ cell lattice through intermediary cells initialized to $|0\RR$.}
\end{center}
\end{figure}

The state could just as well have begun in a superposition state $|\phi\RR_{0}=\alpha|0\RR_{0}+\beta|1\RR_{0}$, in which case the BQCA sequence will transport the state to site $n-1$ as shown in Fig. \ref{fig:1}.  It does so by first mapping the product state into a two particle entangled state, then shifting the adjacent $1$'s two cells per update and finally mapping back to a product state at the other boundary.  The entanglement present during transport is evident from the space time diagrams in Fig. \ref{fig:1}.  The first diagram shows the probability density for each spin to be in state $|1\RR$ defined by $P_{1}(\rho_{j})=Tr[|1\RR_{j}{_{j}\LL}1|\rho_{j}]$, where $\rho_{i}=Tr_{not\  i}(\rho)$ is the reduced state of the spin at site $i$ of the global state $\rho$.  The second diagram displays the reduced von Neumann entropy, defined by $S(\rho_{i})=-Tr(\rho_{i}\log\rho_{i})$.  Starting from a pure separable state and evolving unitarily, the residual mixedness of each spin in the adjacent pair results from mutual entanglement.  We explore in more detail the dynamics of entangled states below.  By linearity, using the above BQCA sequence,  any state $\rho_{0}$ can be swapped through $n-1$ lattice sites prepared in $|0\RR$ in a time $T_{n}\simeq n\pi/(4g)$.  At the cost of one additional update on the $B$ species, the states of two qubits on the ends of a chain can be swapped via the sequence:  $\sigma_{z}^{0}\sigma_{z}^{n-1}M^{B}({\bf 1},e^{-i\frac{\pi}{2}\sigma_{x}},e^{-i\frac{\pi}{2}\sigma_{x}},e^{-i\pi\sigma_{x}})$ $[ M({\bf 1},e^{-i\frac{\pi}{2}\sigma_{x}},e^{-i\frac{\pi}{2}\sigma_{x}},e^{-i\pi\sigma_{x}})] ^{n/2}$, as shown in Fig. \ref{fig:2}.

The QCA state transport time $T_{n}=n\pi/(4 g)$ is provably \cite{Khaneja:Udecomp} the minimal time to translate a quantum operator over $n-1$ sites of a spin chain with pairwise interactions.  This protocol should be compared to the total propagation time using ``soliton operators" proposed in \cite{Khaneja:Infotrans}.  There a single qubit state is encoded into three qubits which then propagate through the chain and are decoded to one qubit at the end.  Similar to the present proposal, the discrete evolution is generated via the Ising interaction punctuated by homogeneous single bit gates.  The swapping time using the soliton operators is $T_{sol\ n}=(n+1)\pi/(4g)$, slightly longer than the present method.  Both pulse sequences require addressability at the boundaries but the simpler QCA sequence has the additional requirement of an architecture that supports an alternating array of two species.  It should be emphasized that neither of these methods are true swap sequences, in the sense that any quantum information encoded in the intervening cells will be disturbed during the sequence.  They may be useful in quantum architectures where quantum ``memory" is stored in qubits spatially separated from each other by ``bus" qubits initialized to the state $|0\RR$ that act as conduits for quantum information.  Architectures with this kind of sparsely distributed memory avoid correlated errors induced by the environment and can make the system amenable to quantum error correction.    

\begin{figure}
\begin{center}
\includegraphics[scale=0.65]{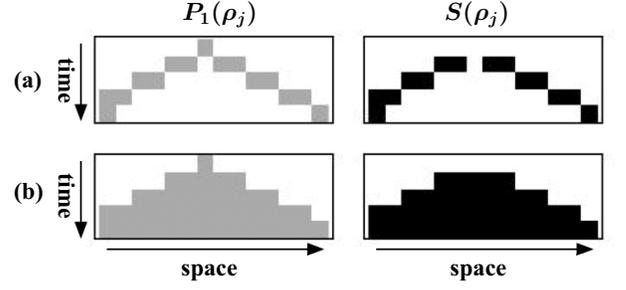}
\caption{\label{fig:3}Generating entangled states beginning with a ``seeded" qubit in the superposition state $1/\sqrt{2}(|0\RR+|1\RR)$. (a)  A maximally entangled pair at the boundaries:  $1/\sqrt{2}(|0\RR_{0}\otimes|0\RR_{n-1}+|1\RR_{0}\otimes|1\RR_{n-1})$. (b) The $n$-spin Greenberger-Horne-Zeilinger (GHZ) state:  $1/\sqrt{2}(|0\ldots0\RR+|1\ldots1\RR)$.}
\end{center}
\end{figure}

BQCA rules also can readily be constructed to distribute entanglement.  Consider the creation of an entangled pair of qubits at the boundaries of an open chain of size $n\geq4$.  Choosing boundary conditions $(\Sigma_{L}=0,\Sigma_{R}=0)$, we begin with a single qubit ``seeded" to the superposition state $1/\sqrt{2}(|0\RR+|1\RR)$ near the middle of the chain with all other spins initialized to the state $|0\RR$ and apply a QCA sequence to create the maximally entangled pair described by the state $1/\sqrt{2}(|0\RR_{0}\otimes|0\RR_{n-1}+|1\RR_{0}\otimes|1\RR_{n-1})$.  The particular BQCA sequence and optimal location of the seed spin will depend on the size of the lattice $n$.  We choose a convention that this spin be of the $A$ species and closest to the middle of the chain.  For $n=4k$, $k\in {\bf N}$, the seed spin is located at site $n/2$ and the update sequence is:  $e^{-i\frac{\pi}{4}\sigma_{z}^{0}}M^{B}({\bf 1},e^{-i\frac{\pi}{2}\sigma_{x}},e^{-i\frac{\pi}{2}\sigma_{x}},e^{-i\pi\sigma_{x}})$ $[M({\bf 1},e^{-i\frac{\pi}{2}\sigma_{x}},e^{-i\frac{\pi}{2}\sigma_{x}},e^{-i\pi\sigma_{x}})]^{k-1}$ $M({\bf 1},e^{-i\frac{\pi}{2}\sigma_{x}},e^{-i\frac{\pi}{2}\sigma_{x}},e^{-i\frac{\pi}{2}\sigma_{x}})$.  Similarly, for $n=4k+2$, the seeded spin is located at site $n/2-1$ and the update sequence is:  $e^{-i\frac{\pi}{4}\sigma_{z}^{0}}[M({\bf 1},e^{-i\frac{\pi}{2}\sigma_{x}},e^{-i\frac{\pi}{2}\sigma_{x}},e^{-i\pi\sigma_{x}})]^{k}$ $M({\bf 1},e^{-i\frac{\pi}{2}\sigma_{x}},e^{-i\frac{\pi}{2}\sigma_{x}},e^{-i\frac{\pi}{2}\sigma_{x}})$.  The sequence works by updating the state $|0\ldots010\ldots0\RR$ once with the rule $M({\bf 1},e^{-i\frac{\pi}{2}\sigma_{x}},e^{-i\frac{\pi}{2}\sigma_{x}},e^{-i\frac{\pi}{2}\sigma_{x}})$, creating a separated pair of adjacent spins in the state $|1\RR$.  These pairs then propagate outward under the same rule used to transport quantum information to the boundaries.   An example for $n=14$ is shown in Fig. \ref{fig:3}(a).  The total time to produce an entangled pair at the boundaries is calculated,
using Eqs.~\ref{Mtot},\ref{M1} and again assuming $t_{s}$ is negligible, to be $T_{n}=(4+n/2)\pi/(4g)$.  By a similar argument to the optimality of quantum state transport sequence, the present sequence for distributing entanglement is optimal within the QCA framework.  However, if one is allowed to perform measurements as well as unitary evolution, then an entangled pair can be produced at the ends using ``entanglement swapping".  Given the same architecture under present consideration, it has been shown \cite{Brennen:Qarch} that a maximally entangled pair can be swapped to the ends of a spin chain in a time, $T=\pi/(2g)$ {\it independent} of the length.  The entangled pair could then be used as a resource to teleport a quantum state from one end of the chain to the other.  Naturally, any protocol to deterministically distribute quantum states must preserve causality and is fundamentally limited by the speed of light which enters into the protocol through classical processing of measurement results over the length of the chain. 
          
Multi-particle entanglement can be constructed using a slight variation of the sequence for distributing entangled pairs.  As above we assume boundary conditions $(\Sigma_{L}=0,\Sigma_{R}=0)$ with a seeded qubit in the superposition $1/\sqrt{2}(|0\RR+|1\RR)$.  An $n$-spin GHZ state $1/\sqrt{2}(|0\ldots0\RR+|1\ldots1\RR)$ can be generated as follows.  For $n=4k$, $k\in {\bf N}$, the seed spin is located at site $n/2$ and the update sequence is:  $e^{-i(-1)^{k}\frac{\pi}{4}\sigma_{z}^{0}}$ $[M({\bf 1},e^{-i\frac{\pi}{2}\sigma_{x}},e^{-i\frac{\pi}{2}\sigma_{x}},e^{-i\pi\sigma_{x}})]^{k}$.  Similarly, for $n=4k+2$, the ``seeded" spin is located at site $n/2-1$ and the update sequence is:  $e^{i(-1)^{k}\frac{\pi}{4}\sigma_{z}^{0}}M^{B}({\bf 1},e^{-i\frac{\pi}{2}\sigma_{x}},e^{-i\frac{\pi}{2}\sigma_{x}},e^{-i\pi\sigma_{x}})$ $[M({\bf 1},e^{-i\frac{\pi}{2}\sigma_{x}},e^{-i\frac{\pi}{2}\sigma_{x}},e^{-i\pi\sigma_{x}})]^{k}$.  An example for $n=14$ is shown in Fig. \ref{fig:3}(b).  The total time for the BQCA sequence is:  $T_{n}=n\pi/(8g)$.

Choosing other, more complicated, initial states can allow BQCA generation of many classes of multi-particle entanglement.  For example, if we fix boundaries at $(\Sigma_{L}=0,\Sigma_{R}=0)$ and initialize the $n$ cell lattice to $|\Psi=e^{-i\frac{\pi}{4}\sum_{j=0}^{n-1}\sigma_{y}^{j}}|0\dots0\RR$, then the resultant state after one update by the rule:  $M({\bf 1},e^{-i\frac{\pi}{4}\sigma_{z}},e^{-i\frac{\pi}{4}\sigma_{z}},e^{-i\frac{\pi}{2}\sigma_{z}})$ is characterized by many-particle quantum correlations.  In fact it is equivalent, up to local unitaries, to the so called cluster state \cite{Briegel:Cluster}:
\be
|\Psi_{n\ clus}\RR=\frac{1}{2^{n/2}}\bigotimes_{a=0}^{n-1}(|0\RR_{a}\sigma_{z}^{a+1}+|1\RR_{a}),\;
\ee
with the convention $\sigma_{n}={\bf 1}$.
These states obtain maximal reduced entropy of every spin.  Most notably, they have the property of maintaining persistency of entanglement between the remaining set of qubits when some are lost (depolarized, measured, etc.).  They have exponentially large Schmidt number, namely any expansion of the state in terms of a product basis will require at least $2^{n/2}$ terms.  This is to be contrasted with the $n$-spin GHZ state which has Schmidt number 2. 

\section{Entanglement Dynamics}
\subsection{Quantifying Multi-Spin Entanglement}
Generally, QCA evolution can take a configuration of spins prepared in a product state to a number of different entangled states. In order to characterize the dynamics of entanglement, it would be helpful to have a single parameter that quantifies the amount of multi-particle entanglement contained in a state at any given time step.  A good measure of entanglement should capture the nonlocal nature of the quantum correlations of the spins and therefore should be a function on the state that is non increasing, on average, under local operations and classical communication.   Because entanglement can be shared in different ways by different subsets (parties) of the spins in the lattice, this is no single functions that describes multi-partite entanglement.  For the purposes of this paper we quantify the amount of multi-spin entanglement with a function on pure states of $n$ qubits introduced in \cite{Meyer:entmeas} and expressible as:
\be
R(|\psi\RR)=2\Biggl(1-1/n\sum_{j=0}^{n-1}Tr[\rho_{j}^{2}]\Biggr).\;
\label{R}
\ee
The measure $R$ is linearly related to the purity of the single qubits averaged over the lattice and satisfies two important properties.  First, $0\leq R(|\psi\RR)\leq 1$, where $R(|\psi\RR)=0$ iff $|\psi\RR$ is a product state, and $R(|\psi\RR)=1$ for some entangled states.  Second, $R(\psi\RR)$ is invariant under local unitaries $U_{j}$.

A significant advantage of this function over other possible measures is that it can be observed in a straightforward manner by measurement.  This can be done by introducing a second, identical 1D lattice and interacting the two lattices, bitwise, with a third addressable 1D lattice that can be prepared and measured.  The measurement requires only that each lattice be addressable but does not require addressability of cells within the lattice and is described in \cite{Brennen:Entmeas}.  A deficiency of $R$ as an entanglement measure is that it cannot distinguish sub-global entanglement.  For example, in a $n=4$ lattice, the product state of two maximally entangled Bell states and the 4-spin GHZ state both have $R$ values equal to 1.  This should be kept in mind when quantifying the entangling capacity of BQCA rules as is done below.  In principle, there are other measurements that can be carried out over a many spin system to distinguish one type of shared entanglement from another.   

\subsection{Generating multi-spin entanglement}
In Sec. III, we considered some examples of BQCA rules that generate and distribute entangled states.  In this section, instead of searching for BQCA sequences that generate particular entangled states, we explore some basic properties of the rules themselves by way of two examples.  While these examples are not intended to simulate any particular physical system, they do illustrate some universal behaviors of BQCA and indicate the computational power of simple rules applied over local, in this case 3 cell, neighborhoods.  

It would be beneficial if some predictive statements could be made about the behavior of QCA.  We know that classical cellular automata have the property that a globally reversible rules follow closed evolution \cite{Wolfram:NKS}.  That is, any initial configuration will evolve back to itself after a characteristic period that depends on the rule and the configuration.  The maximum period of evolution is the size of the configuration space, which for an $n$ cell lattice with 2 states per cell is $2^{n}$.  Linear CA rules are those that satisfy the property that for an initial configuration that is a mixture of two configurations, $\vec{u}=a\vec{v}+b\vec{w}$, where $a,b\in{\bf R}$,  the rule acts linearly on the inputs:  $M\vec{u}=aM\vec{v}+bM\vec{w}$.  If the periods of the configurations  $\vec{v}$ and $\vec{w}$ under rule $M$ are $T_{M}(\vec{v})$ and $T_{M}(\vec{w})$ respectively, then the period of $\vec{u}$ is $T_{M}(\vec{u})=lcm(T_{M}(\vec{v}),T_{M}(\vec{w}))$.  For an arbritary mixture of $m$ configurations, $\vec{u}=\sum_{k=1}^{m}a_{k}\vec{v_{k}}$, the period is $T_{M}(\vec{u})=lcm(\{T_{M}(\vec{v_{k}}\})$.

\begin{figure}
\begin{center}
\includegraphics[scale=0.5]{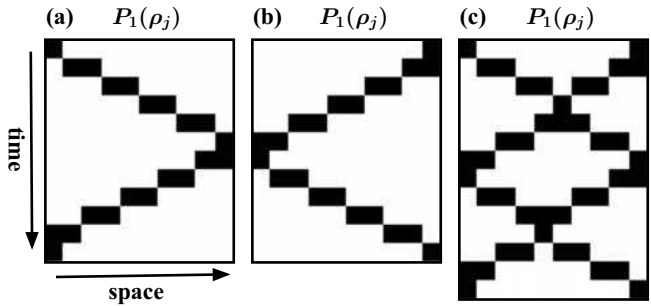}
\caption{\label{fig:4}Evolution of three computational basis states by the rule $M({\bf 1},e^{-i\frac{\pi}{2}\sigma_{x}},e^{-i\frac{\pi}{2}\sigma_{x}},e^{-i\frac{\pi}{2}\sigma_{x}})$ over an $n=10$ lattice with boundaries fixed at $|0\RR$.  Shown are the probability density space time diagrams over one period for the initial states:  (a) $|10\ldots00\RR$, (b) $|0\ldots01\RR$, and (c) $|10\ldots01\RR$.  The evolution of the state $|0\ldots0\RR$ is trivial.  At no time is entanglement generated.}
\end{center}
\end{figure}

\begin{figure}
\begin{center}
\includegraphics[scale=0.6]{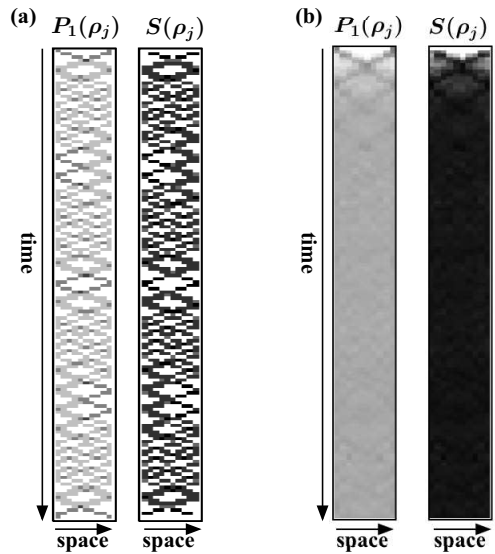}
\caption{\label{fig:5}Entanglement dynamics visualized by the space time histories of the evolution of a chain of $10$ spins by two BQCA rules.  The boundaries are fixed at $|0\RR$ and the initial state is the same for both rules with all qubits initialized to $0\RR$ except for the qubits at sites $0$ and $n-1$ each in the superposition state $1/\sqrt{2}(|0\RR+|1\RR)$.  (a) Rule $M({\bf 1},e^{-i\frac{\pi}{2}\sigma_{x}},e^{-i\frac{\pi}{2}\sigma_{x}},e^{-i\frac{\pi}{2}\sigma_{x}})$.  (b)  Rule $M({\bf 1},e^{-i\frac{\pi}{4}\sigma_{x}},e^{-i\frac{\pi}{4}\sigma_{x}},e^{-i\frac{\pi}{4}\sigma_{x}})$.}
\end{center}
\end{figure}

\begin{figure}
\begin{center}
\includegraphics[scale=0.55]{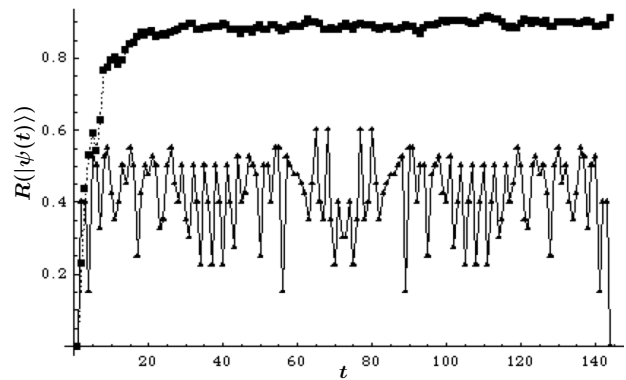}
\caption{\label{fig:6}Multi-qubit entanglement generated during the BQCA evolution plotted in Fig.~\ref{fig:5}.  Entanglement of the global state is plotted for the rules $M({\bf 1},e^{-i\frac{\pi}{2}\sigma_{x}},e^{-i\frac{\pi}{2}\sigma_{x}},e^{-i\frac{\pi}{2}\sigma_{x}})$ (triangles) and  $M({\bf 1},e^{-i\frac{\pi}{4}\sigma_{x}},e^{-i\frac{\pi}{4}\sigma_{x}},e^{-i\frac{\pi}{4}\sigma_{x}})$ (boxes).}
\end{center}
\end{figure}

We consider the BQCA rule $M1\equiv M({\bf 1},e^{-i\frac{\pi}{2}\sigma_{x}},e^{-i\frac{\pi}{2}\sigma_{x}},e^{-i\frac{\pi}{2}\sigma_{x}})$ acting on an initial product state of spins with boundaries fixed at $(\Sigma_{L}=0,\Sigma_{R}=0)$.  For the computational basis states with $|0\RR$ on the interior qubits and $|0\RR$ or $|1\RR$ at the ends, the evolution is reversible as shown in Fig.~\ref{fig:4}.  These four initial states have characteristic periods $1,11,$ and $13$ and at no time is entanglement generated by the rule.  If the initial state is chosen as an evenly weighted superposition of these four basis states: $|\psi(0)\RR=e^{-i\frac{\pi}{4}(\sigma_{y}^{0}+\sigma_{y}^{n-1})}|0\ldots0\RR$, then entanglement is generated by the rule because non-separable phases accumulate on the co-evolving basis states.  The space time diagrams of probability density and reduced entropy are shown in Fig.~\ref{fig:5}a.  The evolution is periodic with a period given by $T=lcm(1,11,13)=143$.  The multi-qubit entanglement during the evolution is plotted in Fig.~\ref{fig:6}.  The entanglement never attains values above $R(|\psi(t)\RR)=0.6$, and this is evident in the space time plot of reduced entropy which shows that at any given time step no more that $8$ out of $10$ spins are entangled with each other.

The entanglement dynamics are dramatically different for the same system and initial state evolving under the BQCA rule $M2\equiv M({\bf 1},e^{-i\frac{\pi}{4}\sigma_{x}},e^{-i\frac{\pi}{4}\sigma_{x}},e^{-i\frac{\pi}{4}\sigma_{x}})$.  This rule rotates each spin by half the amount of the rule $M1$, however, the quantum dynamics does not follow a simple composition rule, i.e. $M1\neq (M2)^{2}$.  This is because the underlying Hamiltonians that update the states of species $A$ and $B$ are non-commuting so that correlations build up at rates that are not linearly related.  The space time diagrams of probability density and reduced entropy plotted in Fig.~\ref{fig:5}b show that after three steps, correlations spread throughout the lattice.  No periodicity is evident, and after roughly $20$ steps, the multi-qubit entanglement, plotted in Fig.~\ref{fig:6}, saturates at a value of approximately $R(|\psi(t)\RR)=0.9$ with small fluctuations.  One way to discern whether the rule is generating many different classes of entangled states during the evolution is to examine temporal variation of Schmidt numbers of the state $|\psi(t)\RR$ over the set of all $2^{n-1}$ bipartite divisions of the $n$ lattice qubits.  The Schmidt numbers are invariant under local unitary operations and under a bipartite division of $k$ and $n-k$ qubits, their range is the integers in the interval $[0,min\{2^{k},2^{n-k}\}]$.  We have calculated the history of Schmidt numbers over the evolution period and find that the rule generates a large number of different classes of entangled states.  This demonstrates that the rule $M2$ explores a larger volume of the Hilbert space of pure states in $\mathcal{H}_{2}^{\otimes n}$ than does rule $M1$ for the given initial state. 

It is an open question under what BQCA rules and initial states is the set of states generated during evolution dense on the Hilbert space of pure states in $\mathcal{H}_{2}^{\otimes n}$.  One might expect that rules that rotate the updated spins by an angle that is an irrational multiple of $\pi$ would accomplish this for a large class of initial states. 

\section{Non-unitary Rules}
\subsection{Formulation}
Up to this point we have described how to implement a class of unitary BQCA rules.  In general one would like to have a prescription for implementing {\it non-unitary} rules as well.  Of the 256 Wolfram rules for radius $1$, two state classical CA, only sixteen are locally invertible.  They are given by:
\be
\begin{array}{lll}
U_{j,k,l,m}&=&|00\RL00|\otimes(\sigma_{x})^{j}+ |01\RL01|\otimes(\sigma_{x})^{k}\\
&+&|10\RL10|\otimes(\sigma_{x})^{l}+ |11\RL11|\otimes(\sigma_{x})^{m},\;
\end{array}
\label{unitary}
\ee
where $\{j,k,l,m\}\in\{0,1\}$.  Among classical CA, all the unitary rules generate rather simple behavior compared to the complex dynamics generated from some of the other, non-unitary, rules \cite{Wolfram:NKS}.  For example, rule 110 is described by the following update table:
\begin{center}
  $R^{110}$:
  \begin{tabular}{cccccccc}
   111 & 110 & 101 & 100 & 011 & 010 & 001 & 000 \\ \hline
    0  &  1  &  1  &  0  &  1  &  1  &  1  &  0
  \end{tabular}
 \end{center}
which is not one-to-one because both states $011$ and $001$ map to $011$.  As mentioned above, rule 110 is universal and we might expect that for quantum cellular automata there are interesting dynamics to be explored when the register is no longer a closed system obeying unitary dynamics but interacts with an environment in an irreversible way.

Non-unitary rules correspond to completely positive maps applied to a system dependent on the state of the neighborhood.  A general, completely positive map on a quantum state $\rho$ can be written as a superoperator in the Krauss representation \cite{Krauss} as:  $S(\rho)=\sum_{\mu=1}^{k}F_{\mu}\rho F_{\mu}^{\dagger}$, where the total number of effects $F_{\mu}$ is $k$ and trace preservation of the state is ensured by the constraint $\sum_{\mu=1}^{k}F_{\mu}^{\dagger}F_{\mu}={\bf 1}$.  In the QCA context, the effects acting on a three cell neighborhood $(j-1,j,j+1)$ are a sum of actions on qubit $j$ induced by orthogonal states of the qubits at sites $j-1$ and $j+1$.  The superoperator can be written as the composition
\be
S_{j}(\rho)=S_{j}^{00}\circ S_{j}^{01}\circ S_{j}^{10}\circ S_{j}^{11}(\rho),\;
\label{decomp}
\ee     
where
\be
S_{j}^{ab}(\rho)=|ab\RL ab|\otimes\sum_{\mu=1}^{k_{ab}} f_{\mu}^{ab}\rho f_{\mu}^{ab \dagger}\otimes|ab\RL ab|.\;
\label{singleS}
\ee
Here, $k_{ab}$ denotes the number of effects that act on the updated qubit $j$ when the neighborhood is in the state $|ab\RR_{j-1,j+1}$.  The single qubit superoperators are trace preserving, i.e. $\sum_{\mu=1}^{k_{ab}} f_{\mu}^{ab \dagger}f_{\mu}^{ab}={\bf 1}$.
As with the unitary maps, the maps $S_{j}$ and $S_{j+2}$ commute, so qubits at every other site can be updated in parallel.  We denote a total BQCA update sequence from time $t$ to $t+1$ by: $\rho(t+1)=\$(\rho(t))=\$^{A}\circ\$^{B}(\rho(t))$, where
\be
\begin{array}{lll}
\$^{A}(\rho)&=&S_{0}\circ S_{2} \circ \cdots\circ S_{n-2}(\rho),\\
\$^{B}(\rho)&=&S_{1}\circ S_{3} \circ \cdots\circ S_{n-1}(\rho).\;
\end{array}
\ee

As an example, the CA rule 110 updating the state of a {\it qubit} at site $j$ is written
\be
R^{110}_{j}(\rho)=F_{1}(j)\rho F_{1}^{\dagger}(j)+F_{2}(j)\rho F_{2}^{\dagger}(j),\;
\ee
where
\be
\begin{array}{lll}
F_{1}^{j}&=&|00\RL00|\otimes{\bf 1}^{j}+ |10\RL10|\otimes{\bf 1}^{j}\\
& &+ |11\RL11|\otimes\sigma_{x}^{j}+|01\RL01|\otimes|1\RR_{j}{_{j}\LL}1|,\\
F_{2}^{j}&=& |01\RL01|\otimes|1\RR_{j}{_{j}\LL}0|.\;
\end{array}
\ee
The rule can be decomposed into unitary and non-unitary BQCA rules as:
\be
R^{110}_{j}(\rho)=S_{j}^{01}(M({\bf 1},{\bf 1},{\bf 1},\sigma_{x}^{j})\rho M({\bf 1},{\bf 1},{\bf 1},\sigma_{x}^{j})),\;
\ee
where
\be
\begin{array}{lll}
S_{j}^{01}(\rho)&=&|01\RL 01|\otimes(|1\RR_{j}{_{j}\LL}1|\rho|1\RR_{j}{_{j}\LL}1|\\
&+&|1\RR_{j}{_{j}\LL}0|\rho|0\RR_{j}{_{j}\LL}1|)\otimes|01\RL 01|).\;
\end{array}
\ee
When the neighborhood is in state $|01\RR$, rule 110 has the effect of an amplitude damping channel on qubit $j$, i.e. it maps a mixed state to a pure state $|1\RR$.
\subsection{Simulation}
In this section we demonstrate how to implement non-unitary rules within a QCA architecture.  A non-unitary map on a quantum system residing in a Hilbert space ${\cal H}^{s}$ can be thought of as open system dynamics that arise from unitary operation in the combined space ${\cal H}^{s}\otimes{\cal H}^{e}$ of the system and some environment, followed by tracing over the environmental degrees of freedom.  Any superoperator on a system of dimension $d$ can be realized in this fashion with an environment of dimension at most $d^{2}$; meaning that the maximal number of effects in a superoperator expansion is $k=d^{2}$ \cite{Davies:Qmeas}.  

Generally, implementing control over a combined system and environment of this size is difficult, however, it has been shown that by using measurement and feedback, a single qubit environment is sufficient to simulate open system dynamics \cite{Lloyd:Opsysdyn}.  The simulation works by coupling a single qubit, $e$, prepared in the state $|+\RR_{e}=1/\sqrt{2}(|0\RR_{e}+|1\RR_{e})$ to the system, $s$, via a Hamiltonian $H_{sym}=\gamma P\otimes\sigma_{z}$, where $P$ is a projector onto a pure state in $s$.  The corresponding unitary operation is $U(t)=e^{-iH_{sym}t}=\cos(\gamma tP)\otimes{\bf 1}-i\sin(\gamma t P)\otimes\sigma_{z}$.  By suitable averaging techniques, namely conjugating short time evolution, $U(\Delta t)$, where $\Delta t\ll t$, with unitary operations on $s$, the projector $P$ can be transformed into any positive, unit trace, operator $\bar{P}$.  Finally, $e$ is measured in the $\sigma_{x}$ basis and the result is fedback to $s$ with one of two unitaries  $U_{0},U_{1}$, depending on the measurement result.  This process will implement any superoperator described by two effects:  $F_{0}=U_{0}\cos(\gamma t \bar{P})$ and $F_{1}=U_{1}\sin(\gamma t \bar{P})$.  Maps with more than two effects can be simulated by repeated cycles of measurement and feedback.

In the QCA context, we want to activate non-unitary dynamics on a qubit $s$ dependent on the neighborhood state.  This can be accomplished by first ``turning on" a control qubit $c$ dependent on the neighborhood state, then implementing the unitary, $U_{c-sym}=e^{-i |1\RR_{c}{_{c}\LL} 1|\otimes H_{sym}t}$ during the simulation steps described above.  For instance, the superoperator $S_{s}^{01}(\rho)$ can be implemented by first entangling the register and the control qubit (initialized to $|0\RR_{c}$) with the unitary BQCA rule $M^{c}({\bf 1},{\bf 1},e^{-i\frac{\pi}{2}\sigma_{x}},{\bf 1})$.  We choose the projector in $H_{sym}$ to be $P=|1\RR_{s}{_{s}\LL 1|}$ and simulate open system dynamics on $s$ by evolving the combined system $(c,e,s)$ with the unitary $U_{c-sym}=e^{-i\frac{\gamma t}{4}\sigma_{z}^{c}\otimes\sigma_{z}^{e}\otimes\sigma_{z}^{s}}e^{i\frac{\gamma t}{4}\sigma_{z}^{e}\otimes\sigma_{z}^{s}}e^{i\frac{\gamma t}{4}\sigma_{z}^{c}\otimes\sigma_{z}^{e}}$.  The gate $U_{c-sym}$ can be efficiently implemented using pairwise interactions between system pairs $(c,e)$ and $(e,s)$ \cite{Brennen:Entmeas}.  At the measurement stage, the classical result of the measurement on $e$ should be ignored, equivalent to tracing over the environmental degrees of freedom.  This can be accomidated by an environmental qubit that sinks information of the measurement result to a large reservoir (such as an atom that emits spontaneous radiation).  The conditional feedback can be implemented without knowledge of the measurement result using a controlled-unitary operation between $e$ and $s$, namely:  $U_{fb}=|0\RR_{e}{_{e}\LL 0|}\otimes U_{0}^{s}+|1\RR_{e}{_{e}\LL 1|}\otimes U_{1}^{s}$.  Finally, the control qubit $c$ needs to be disentangled from the register with the rule $M^{c}({\bf 1},{\bf 1},e^{i\frac{\pi}{2}\sigma_{x}},{\bf 1})$, and the ``environment" qubit reset to $|+\RR_{e}$.  

Note that this protocol has the unwanted effect of applying a unitary $U_{0}$ to the qubit $s$ regardless of the state of the neighborhood, because the feedback is only conditioned on the environmental state which is initialized to $|0\RR$.  This can be obviated by instead using the feedback gate:  $U_{fb}^{\prime}=|0\RR_{e}{_{e}\LL 0|}\otimes {\bf 1}^{s}+|1\RR_{e}{_{e}\LL 1|}\otimes U_{0}^{\dagger s}U_{1}^{s}$, and after the control has been disentangled from $s$, applying a unitary BQCA on $s$.  For example, after the simulation of $S_{j}^{01}$ with the gate $U_{fb}^{\prime}$, the unitary BQCA rule $M^{s}({\bf 1},U_{0},{\bf 1},{\bf 1})$ will apply the necessary feedback.  By the superoperator decomposition, Eq.\ref{decomp}, any non-unitary BQCA rule can be implemented by at most four instances of the above open systems simulation using the appropriate unitary rule $M^{c}$ for each neighborhood value $|ab\RR$.  

\begin{figure}
\begin{center}
\includegraphics[scale=0.39]{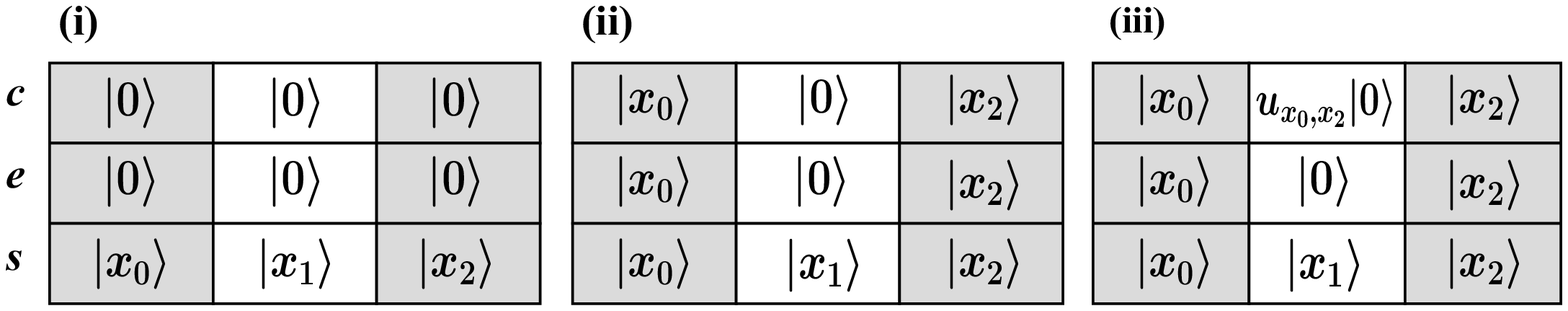}
\caption{\label{fig:7}Sequence of steps to implement non-unitary BQCA rules on a 1D lattice, $s$, using qubits in a control lattice $c$ and an ``environment" lattice $e$.  The lattice cells are assumed addressable along the vertical but not the horizontal direction with grey(white) cells corresponding to qubit species $A(B)$.  Shown is a sequence sketched over a three cell section of the lattices for implementing a non-unitary rule on $B$ species qubits dependent on the $A$ species neighbors.  (i). A product state in the computational basis of $s$ is shown with the lattices $c$ and $s$ initialized to $|0\RR$.  (ii). The state of the $A$ species qubits is fanned out to corresponding sites in $e$ and $c$ using CNOT gates.  (iii). The unitary BQCA rule $M^{B}(u_{00},u_{01},u_{10},u_{11})$ acts on $c$ to activate the controls dependent on the neighborhood.  Non-unitary evolution on $s$ is simulated using interactions between the lattices, measurement on $e$ and feedback on $s$.  Afterward, the inverse of steps (ii-iii) disentangles the three lattices.}
\end{center}
\end{figure}

A possible implementation of non-unitary BQCA rules is shown in Fig. \ref{fig:7} for an architecture with three stacked 1D lattices.  The protocol for interacting registers in an optical lattice is outlined in \cite{Brennen:Entmeas}.      

\subsection{Results}
\begin{figure}
\begin{center}
\includegraphics[scale=0.45]{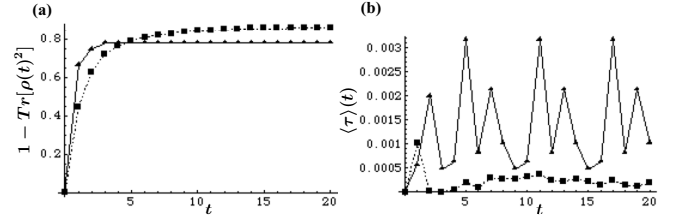}
\caption{\label{fig:8}System evolution by a mixture of unitary and non-unitary rules:  $S^{01}(\rho,p)$.  Results are shown for $p=1/2$ (boxes) and $p=1$ (triangles) with connecting lines to guide the eye.  (a) Mixedness of the system $\rho(t)$.  (b) Entanglement, quantified as the average tangle $\tau$ over all spin pairs.  Increased coupling to the environment increases the entanglement.}
\end{center}
\end{figure}

We investigate the effect of adding decoherence to a unitary BQCA by mixing the  rules $R^{110}$ and $R^{108}=M({\bf 1},{\bf 1},{\bf 1},\sigma_{x}^{j})$.  This is described by a one parameter map on the neighborhood $(j-1,j,j+1)$ written as:
\be
S_{j}(\rho,p)=S_{j}^{01}(M({\bf 1},{\bf 1},{\bf 1},\sigma_{x}^{j})\rho M({\bf 1},{\bf 1},{\bf 1},\sigma_{x}^{j}),p),\;
\ee  
where
\be
\begin{array}{lll}
S_{j}^{01}(\rho,p)&=&|01\RL 01|\otimes((|1\RR_{j}{_{j}\LL}1|\\
&+&\sqrt{1-p}|0\RR_{j}{_{j}\LL}0|)\rho(|1\RR_{j}{_{j}\LL}1|+\sqrt{1-p}|0\RR_{j}{_{j}\LL}0|)\\
&+&p|1\RR_{j}{_{j}\LL}0|\rho|0\RR_{j}{_{j}\LL}1|)\otimes|01\RL 01|).\;
\end{array}
\label{nonurule}
\ee
In the case $p\rightarrow0(1)$, the rule approaches $R^{108}(R^{110})$.  We study the evolution of entanglement under this map when the initial state is chosen to be the superposition of all computational basis states:  $|\psi(0)\RR=e^{-i\frac{\pi}{4}\sum_{j}^{n-1}\sigma_{y}^{j}}|0\RR$.  Note that $R^{108}$ permutes computational basis states so that $|\psi(0)\RR$ is an eigenstate of this rule and no entanglement between spins is generated \footnote{Because $\sigma_{x} \notin $ SU(2), rule 108 is not strictly within the class of implementable BQCA rules in 1D.  However one can use the rule $M({\bf 1},{\bf 1},{\bf 1},e^{i\frac{\pi}{2}\sigma_{x}})$ instead and correct for the phase using a controlled phase gate in the non-unitary implementation stage.}.  However, when the system is subjected to non-unitary rules, entanglement can develop.  This behavior is illustrated in Fig.~\ref{fig:8} for an $n=6$ lattice with boundaries fixed at $(\Sigma_{L}=0,\Sigma_{R}=0)$.  Two global quantities of the spin chain are plotted; the mixedness of the state, $1-Tr[\rho^{2}]$, and the entanglement.  The entanglement over the multi-partite mixed state $\rho$ is calculated by averaging the pairwise tangle $\tau_{ij}$ over all spin pairs $(i,j)$.  The tangle \cite{Wootters:2qubitent} is a monotonic function on pure or mixed states of two qubits assuming the value $0$ for separable states and $1$ for maximally entangled Bell states.  It is defined as a function of the reduced state $\rho_{ij}$ of qubits $i$ and $j$:
$\tau_{ij}=[max\{\lambda_{1}-\lambda_{2}-\lambda_{3}-\lambda_{4}\}]^{2}$
where the $\lambda_{k}$ are the square roots of the eigenvalues, in decreasing order, of
$\rho_{ij}\tilde{\rho}_{ij}$.  Here, $\tilde{\rho}_{ij}$ is the spin-flipped version of $\rho_{ij}$: 
$\tilde{\rho}_{ij}=\sigma_{y} \otimes \sigma_{y} \rho^{*}_{12} \sigma_{y} \otimes \sigma_{y}$.

As the amount of coupling to the environment, quantified by $p$, increases, the amount of entanglement in the spin chain increases.  Additionally, beyond a certain time, the mixedness decreases with increasing coupling.  This latter fact is because rule 110 acts as a neighborhood dependent amplitude damping channel that decreases the reduced entropy of each spin state.  In fact, for $p=1$ the state of the system relaxes after a time $t=4$ to dynamics with period 6 and constant mixedness.  For the mixed rule case, $p=1/2$, in contrast, the system is not driven to periodic evolution but on long time scales experiences small fluctuation in mixedness and entanglement.  

The development of entanglement during open system dynamics, absent in the closed system dynamics can be attributed to the fact that the non-unitary rule acts to dampen the state of a qubit only if the neighborhood is in the state $|01\RR$.  The implementation of this rule requires a three body interaction followed by single spin decoherence.  The net effect is to project the state to one that has some fraction of entanglement.  This is an example of environment assisted entanglement preparation.  Another, well known, example of such a phenomenon is the relaxation of two independent radiating dipoles into a maximally entangled sub-radiant state \cite{Dicke:Supsubrad} when the dipoles are close enough together to see the same electromagnetic field.  The results here show that entanglement can develop even when the environment acts only on one member of a neighborhood of spins.  

It is possible that this effect could be measured in the laboratory as a signature of neighbor dependent environmental coupling.  For instance, consider an array of three atoms of two species $A$ and $B$ trapped inside a high Q cavity with order $ABA$ and aligned perpendicular to the cavity axis.  The species are assumed to have a distinguishable set of two ground state manifolds $|0\RR_{A,B}$ and $|1\RR_{A,B}$ with different resonant excitation frequencies $\hbar\omega_{0\ A(B)}=E_{eA(B)}-E_{0A(B)}$ and $\hbar\omega_{1A(B)}=E_{eA(B)}-E_{1A(B)}$ and possibly different decay rates $\gamma_{A,B}$.  If a laser field at frequency $\omega_{L}$ illuminates all three atoms, dipoles will be excited with dominant dipole-dipole interactions $V_{dd}$ acting pairwise.  Assume that the $\omega_{L}$ is extremely far off resonant to the $|0\RR_{A,B}\ra|e\RR_{A,B}$ transition so that dipoles are excited only when atoms of both species are in state $|1\RR$.  The interactions will shift the energy levels of the two species so that the effective detunings of the field will be $\Delta_{B}\simeq\omega_{L}-\omega_{1\ B}-2 V_{dd}/\hbar$, and $\Delta_{A}\simeq\omega_{L}-\omega_{1\ A}-V_{dd}/\hbar$.  For $V_{dd}$ large enough and appropriate choice of laser frequency, $|\Delta_{B}|\leq\gamma_{B}$ while $|\Delta_{A}|>\gamma_{A}$ so that the field is in resonance with the excited state of atom $B$ but not for the $A$ species.  If the resonant cavity frequency is close to $\omega_{0B}$, then the $B$ species atom will preferentially decay to state $|0\RR_{B}$.  This type of decay corresponds to the non-unitary rule $S=\sigma_{x}^{B}S_{B}^{11}(\sigma_{x}^{B}\rho\sigma_{x}^{B},p)\sigma_{x}^{B}$, in analogy to Eq.~\ref{nonurule}.  The strength of $V_{dd}$ will determine the amount of coupling $p$.       
               
\section{Conclusions}
In this paper, we have shown how to construct a universal class of radius 1, two-state
QCA that are block partitioned and particularly suited for implementation
in systems with a naturally endowed lattice type structure.  In Section IIA we introduced a universal
set of BQCA rules in Eq. 1, which is the formal analog of the subset of 16 unitary rules of Wolfram's  256 radius one rules for classical CA.  We demonstrated how these general rules
can be simulated in a spin lattice with Ising interactions in conjunction
with single qubit rotations (applied in parallel across the lattice). This
is an important result showing the physical relevance of the BQCA to
experimental systems.

Another important result is that we have suggested a new approach to quantum
information: use BQCA to explore the raw computational properties of a
physical system, such as the transport of quantum information and the
generation of long-range quantum correlations throughout the system. This
approach should be viewed as complementary to the standard treatment of
quantum information processing centered around the quantum circuit model of
QC. In Section III we presented several specific examples of how BQCA rules
can be chosen to distribute specialized entangled states across the lattice,
for example, to create a $n$-spin GHZ state, shown in Figure 3b. In order
to visualize the development and spread of entanglement, we have introduced
the idea of plotting a space-time diagram of the reduced entropy at each
cell. We then explored more general entanglement dynamics in Section IV,
with a focus on finding rules that are effective at generating multi-qubit
entanglement. For this task, we utilized the multi-qubit entanglement measure
$R(|\psi\rangle)$ related to the average purity of the constituent qubits.

Perhaps our most innovative contribution to the study of QCA is that we have
developed the formalism in Section V that extends BQCA to open quantum
systems, which evolve according to \emph{non-unitary} rules. The BQCA rule is
represented by an appropriate set of Krauss operators acting on the
system density matrix. This more generalized treatment can be thought of as
including the effect of correlated noise in the quantum evolution. From a
practical standpoint, this extension is important for exploring the effect
of a broader class of errors than is typically treated in the theory of
error correction. On a more fundamental level, the non-unitary BQCA can be
used as a testbed for exploring the interplay between quantum and classical
computation--between quantum and classical correlations in a discrete
dynamical system. As a concrete example exploring this notion, in Section V
we presented simulation results of a non-unitary BQCA rule that can be tuned
continuously from a purely open irreversible evolution using rule 110
(implemented in a block partitioned manner) to a purely closed unitary
evolution using the quantum analog of rule 108. Remarkably, we uncovered the
intriguing result that for a particular initial state entanglement generation is optimized by tuning the
mixing parameter to a finite value, which suggests the possibility of
environment assisted entanglement generation.

The are several outstanding issues regarding QCA that warrant future research.  Can QCA be used to simulate complex classical or quantum dynamics?  The study of QLG algorithms demonstrates that there are non-trivial classical dynamics of a single particle can be studied using a lattice of spins.  In these algorithms, the entanglement generated is typically limited to local neighborhoods. It is worth while investigating whether large scale global entanglement generated by QCA can be used to advantage, perhaps to study properties of classical CA with quantum parallelism.  Also the preliminary study of non-unitary QCA here suggests that in some cases open systems dynamics may be a more efficient way to navigate through Hilbert space than purely unitary dynamics.  For practical implementation one wonders how resilient QCA are to decoherence and noise in the control fields.  Work on decoherence free subspaces (DFS) shows that when a group of quantum systems see the same environment that the noise can protected against by careful encoding of the states \cite{Zarnari:DFS}.  Because cellular automata evolve by use of global control fields the requirements for DFS are natural to this architecture.

\acknowledgments

We appreciate helpful discussions with Bei Lok Hu, Stephan Wolfram, Charles Clark, and Marianna Safronova.  This work was supported in part by ARDA/NSA.

\bibliography{QIbib}

\end{document}